\documentclass[aip,reprint]{revtex4-2}
\usepackage{color}
\usepackage{amsmath}
\usepackage{amssymb}
\usepackage{graphicx}

\begin{document}
\title{Intrinsic dynamical shadowing of point vortices and finite time singularities}
\author{Xavier Leoncini}
\affiliation{Aix Marseille Univ, Univ Toulon, CNRS, CPT, Marseille, France}
\author{Perla El Kettani}
\affiliation{Univ Toulon, Aix Marseille Univ, CNRS, CPT, Marseille, France}
\author{Edgardo Ugalde}
\affiliation{Instituto de F\'{\i}sica, Universidad Aut{\'o}noma de San Luis Potos\'{\i},
M{\'e}xico}

\begin{abstract}
The trajectories of point vortices after a division using reverse collapse route are studied. In this setting, an unexpected peculiar phenomenon appears, corresponding to the shadowing of the vortex breakup. This occurs at least for a finite time, when the system evolves from three to five vortices  after a breakup. An analytical study reveals that this observed numerical phenomenon is related to the condition of scale invariance, a condition necessary for the finite time singularity. Further simulations agree with our analytical findings and indicate that shadowing
can occur with more than five vortices, and as well with higher order
singularities. This phenomenon leads to question what a vortex is
actually measuring, triggering speculations if this shadowing property
could be generalized to other Hamiltonian systems with finite-time
singularities.
\end{abstract}
\maketitle

\section{Introduction}

Unveiling singularities in equations describing a physical system
is usually related to the presence of a physical phenomenon. To cite
a few examples one can think of a phase transition in statistical
physics \cite{Yang_Lee52,Itzykson1989}, the formation of a drop
\cite{Eggers95} or for larger scales the formation of a black hole
\cite{Schwarzschild1916}. One peculiar singularity is the one leading to a three point-vortex collapse pointed
out in papers~\cite{Grobli1877,Synge49,Novikov79,Aref1992}.
This singularity has lead to numerous different type of works \cite{Kimura90,Babiano94,LKZ2000,Leoncini2011,Kudela2014,Badin2018,Kallyadan2022,Grotto2024}.
Point vortices have been used in many different situations, for instance
the statistical property of assemblies of vortices in the context
of negative temperatures have been linked to the emergence of large
scale structure \cite{Onsager49,Edwards74} (observed in decaying
turbulence and its inverse cascade \cite{McWilliams84}), and in
an out of equilibrium context for instance in kinetic theory \cite{Robert92,FOUVRY2016,Fouvry2025},
though in these contexts the possibility of allowing a fluctuation
of the number of vortices due to inverse collapse routes was not considered.
A kind of forced singularity was as well introduced to mimic turbulence
using so called punctuated point vortex models \cite{Carnevale91}.

In this paper we consider the possibility of using the singularity
to divide vortices. More explicitly by changing the direction of time,
we study the motion of three and then five vortices which are consecutively
generated by the division of a vortex according to a reverse collapse
route in the spirit of what was initiated in \citet{Leoncini2011}.

In what follows, we first recall briefly the setting of what a point
vortex is, and how a singularity can occur when given conditions are
met, among them scale invariance of the system. Then we present how
the splitting of a vortex is constructed, from a theoretical point
of view and a numerical one, and display some numerical results which
show some kind of shadowing phenomenon when we move from three to
five vortices. Given this, a perturbative analytical study of the
phenomenon is conducted, and results in giving the scale invariance
conditions for the shadowing to apply. And finally, given the previous
result a conclusive test is performed when splitting a vortex in four
vortices and we conclude.

\section{Setting of the problem}

Point vortices are exact solutions of the Euler equation of a two-dimensional
incompressible flow (see for instance the books\cite{Newton2001nvortex,Sokolovskiy2014dynamics}).
In order to obtain their dynamics, we consider the Euler equation
for the vorticity 
\begin{equation}
\frac{\partial\Omega}{\partial t}+\left\{ \Omega,\psi\right\} =0\:,\:\Omega=-\nabla^{2}\psi\:,\label{vorti}
\end{equation}
where $\psi$ is called the stream function, and $\{\cdot,\cdot\}$
denotes the Poisson brackets. Then we look on what conditions a vorticity
field given by a superposition of Dirac functions 
\begin{equation}
{\normalcolor \Omega(\mathbf{r},t)=\sum^{N}_{i=1}{\color{red}{\normalcolor k_{i}}}\delta\left(\mathbf{r}-{\normalcolor {\color{red}{\color{red}{\normalcolor \mathbf{r}_{i}(t}\mathclose{\normalcolor )}}}}\right)\:}\label{eq:vort_dirac}
\end{equation}
is a solution of (\ref{vorti}). In (\ref{eq:vort_dirac}), $k_{i}$
corresponds to the vorticity of a point vortex localized at point
$\boldsymbol{r}_{i}(t)$ on a plane, with Cartesian coordinates
$(x_{i}(t),y_{i}(t))$. In order to be a a solution of the Euler equation~(\ref{vorti})
the superposition of point vortices have to follow an $N$-body Hamiltonian
dynamics \cite{Machioro94} whose Hamiltonian can be written as
\begin{equation}
H=\frac{1}{4\pi}\sum_{i\ne j}k_{i}k_{j}\ln\left\Vert \boldsymbol{r}_{i}-\boldsymbol{r}_{j}\right\Vert \:,\label{Hamilton}
\end{equation}
where $k_{i}y_{i}$ and $x_{i}$ are the canonically conjugate variables
of the Hamiltonian~(\ref{Hamilton}). The dynamics of each vortex
corresponds its advection on the flow generated by the other vortices.

\begin{figure}
\begin{centering}
\includegraphics[width=7cm]{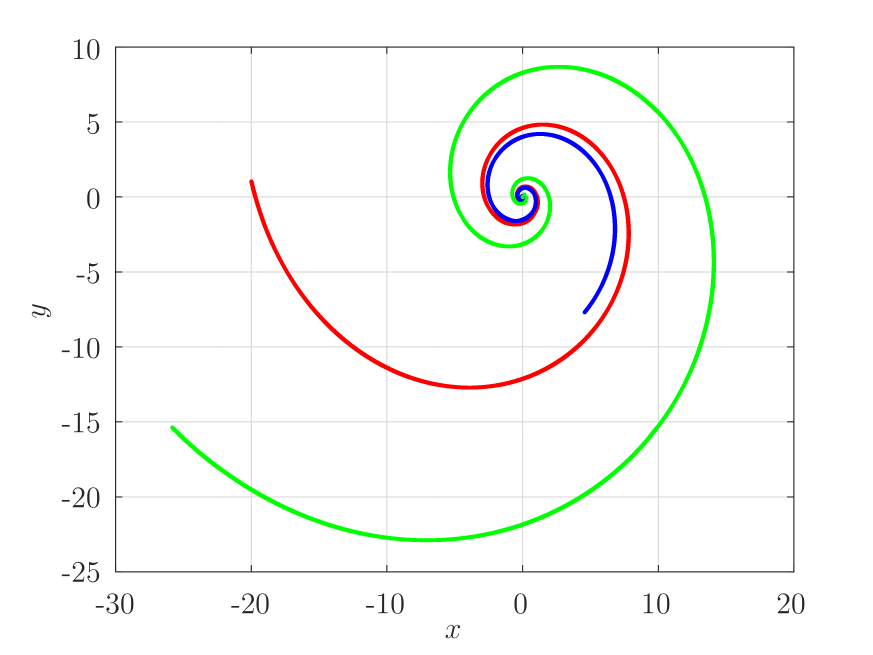}
\par\end{centering}
\caption{Illustration of the collapse in finite time of three point vortices.
Vortex strengths are $k_{1}=-0.5924$ (red), $k_{2}=-0.7358$ (blue)
and $k_{3}=+0.3282$ (green). \label{fig:Votrex_collapse}}
\end{figure}

In order to discuss the dynamics of point vortices, we look at the
continuous symmetries of Hamiltonian (\ref{Hamilton}). We can notice
that it is invariant by translation and by rotation. As a consequence,
the dynamics of system exhibit four integral of the motion, i.e.,
four conserved quantities. We have, the energy corresponding to the
vortex interactions associated to the Hamiltonian, some ``momentum''
conservation (associated to translational invariance, a center of
vorticity, when it can be defined, i.e $\sum k_{i}\ne0$, which will
be the case in the present study)
\begin{equation}
\boldsymbol{C}=\frac{\sum k_{i}\boldsymbol{r}_{i}(t)}{\sum k_{i}}\,,\label{momen1}
\end{equation}
and an ``angular momentum'' associated to the invariance by
rotation 
\begin{equation}
L^{2}=\sum^{N}_{i=1}k_{i}r^{2}_{i}(t).\label{momen2}
\end{equation}
It is convenient to rewrite this condition in order to get rid of
the choice of a specific origin and using the two previous constants
of the motion to generated another one:
\begin{equation}
K=\left[\left(\sum_{i}k_{i}\right)L^{2}-\left(\sum_{i}k_{i}\right)^{2}\boldsymbol{C}^{2}\right]=\sum_{i\ne j}k_{i}k_{j}r^{2}_{ij}\:.\label{eq:ang_momentum-1}
\end{equation}
Only three of these four constants are truly independent, i.e., are
in involution and given the Arnold-Liouville theorem, this implies
that the motion is integrable if the number of point vortices $N$
is less or equal to three, and it is generically chaotic if $N\ge4$
\cite{Novikov78,Aref80}. The global different integrable trajectories
have been studied and characterized in \cite{Aref79}, and the dynamics
with more vortices \cite{Aref80,Dritschel2015} or their use to study
flows leading to chaotic advection have been carried out\cite{Aref84,Boatto99,Laforgia01,LZ02,LKZ03}.
Among the different possible motion, one may pinpoint a striking one
leading to the collapse of three point vortices in a finite time \cite{Synge49,Novikov79,LKZ2000}.
An illustration of this phenomenon is displayed in Fig.~\ref{fig:Votrex_collapse}
.

Let us recall the conditions for this singularity to occur, the first
condition on the vortex strength is usually written as 
\begin{equation}
\sum^{3}_{i=1}\frac{1}{k_{i}}=0\:,\label{eq:collapse_condition_three_vortex}
\end{equation}
and the second condition is more geometrical and linked to the constants
of the motion, indeed the collapse has to be compatible with these
and we must then have
\begin{equation}
K=0\:.\label{eq:ang_momentum_condition}
\end{equation}
In fact the first condition (\ref{eq:collapse_condition_three_vortex})
is intimately related to another property of the Hamiltonian of point
vortices, the scale invariance. Indeed, due to the logarithmic interaction
between vortices, rescaling all length by some constant let (\ref{Hamilton})
unchanged if the condition 
\begin{equation}
\sum_{i\ne j}k_{i}k_{j}=0\:,\label{eq:scale_invariance_gen}
\end{equation}
is met, which is actually the same condition as (\ref{eq:collapse_condition_three_vortex})
for three vortices. When these conditions are fulfilled, the area
of the triangle formed by the three vortices decreases linearly (in
time) and the three vortices merge onto the center of vorticity in
a finite time, as the area of the triangle formed by the vortices
is contracting and rotating, but its shape is unchanged. Last but
not least, and related to the time-reversal symmetry of Hamiltonian
dynamics, a symmetric change of initial conditions can as well lead
to infinite dilatation of the triangle. As a side note, this new continuous
``symmetry'' related to scale invariance does not lead to an extra
constant of the motion, as the Poisson brackets giving rise to the
Hamiltonian dynamics is not invariant by the transformation and needs
as well a rescaling of time.

This time-reversal symmetry lead to the possibility of generating
a distribution of point vortices by successive division of vortices
according to a reverse collapse route. This possibility was explored
in this paper \cite{Leoncini2011}, and it was shown that this vortex generation
was always possible, and leads to some kind of dissipation, while
conserving the total vorticity of the flow. Moreover, the distribution
of vortex strength after a large number of divisions was studied and
exhibited some form of self-similarity which was as well explained.
In this paper we are more focused on what happens dynamically after
a break up, aiming at the spatial evolution of the distribution of
vorticity. For this purpose, in the next section we define in more
details the considered setup.

\section{Physical conditions of vortex splitting and numerical set up}

\subsection{Conditions on vortex strength}

In order to be consistent with what we expect from a collapse, it
appears as natural to keep the total vorticity constant before and
after a vortex splitting, and as well to satisfy the conditions for
a vortex collapse to appear. If we divide a vortex located in a point
$O$, and the three resulting vortices are labeled with running letters
$I,J\in\{A,B,C\}$, we have for a 3-vortex division
\begin{eqnarray}
k_{O} & = & \sum_{I}k_{I}\label{eq:Vorticity conservation}\\
0 & = & \sum_{I\neq J}k_{I}k_{J}\:.\label{eq:Scale_free rule}
\end{eqnarray}
Taking the square of Eq.~(\ref{eq:Vorticity conservation}) and using
Eq.~(\ref{eq:Scale_free rule}) we end up with
\begin{eqnarray}
\sum_{I}k_{I} & = & k_{O}\label{eq:plane}\\
\sum_{I}k^{2}_{I} & = & k^{2}_{O}\;,\label{eq:sphere}
\end{eqnarray}
the values of the $k_{I}$ belong to the circle resulting from the
intersection of the the sphere of radius $|k_{O}|$ and the plane
defined by Eq.~(\ref{eq:plane}), so the choice can be parametrized
by an angle (see \cite{Leoncini2011} for more details).

Let us imagine that after sometime we divide again in three for instance
the vortex $A$ according to the previous rules (replacing this $O$
by $A$, and with running letters $I,J\in\{D,E,F\}$). One can also
verify that the global scale invariance property is preserved when
imposing the local collapse rule~(\ref{eq:Scale_free rule}). Indeed
the three new vortices will add the following terms in the global
term~(\ref{eq:scale_invariance_gen})
\begin{align*}
\sum_{I,\ell\neq A}k_{I}k_{\ell}+\sum_{I\neq J}k_{I}k_{J} & =\sum_{\ell\neq A}k_{\ell}\sum_{I}+\sum_{I,J}k_{I}k_{J}\\
 & =\sum_{\ell\neq A}k_{K}k_{A}+0\:,
\end{align*}
where $\ell\in\{B,C\}$, and this is exactly the missing term in Eq.~(\ref{eq:scale_invariance_gen})
resulting from the splitting of vortex $A$ (we used here the equations~(\ref{eq:Vorticity conservation} and \ref{eq:Scale_free rule}).
And so the new Hamiltonian governing the equations remains scale invariant.

\subsection{Geometrical conditions}

When a vortex located in $A$ is divided, we are creating a triangle
(of zero size), but in order to be consistent with the dynamics we
impose that $A$ is the center of vorticity of the three newly created
vortices
\begin{equation}
\sum_{I}k_{I}\boldsymbol{AI}=\boldsymbol{0}\:,\label{eq:Local_center of vorticty}
\end{equation}
and that this does not affect either the angular momentum, hence we
impose as well 
\begin{equation}
\sum_{I\ne I}k_{I}k_{J}IJ^{2}=0\:.\label{eq:local_angular_momentum_condition}
\end{equation}
Since at the moment of splitting all points $I$ coincide with point
$A$, it is obvious that given the condition~(\ref{eq:Vorticity conservation}),
after a given number of division, the total center of vorticity is
unchanged and preserved, as well as the global constant of the motion~(\ref{eq:ang_momentum-1}).

The condition (\ref{eq:local_angular_momentum_condition}) implies
some conditions on the shape of triangle once the vortex strength
have been fixed. Since this condition is invariant by rotation, and
the system is scale free, let us investigate what it imposes using
three points $C=(0,0)$, $B=(1,0)$ and $A=(x,y)$. The condition
(\ref{eq:local_angular_momentum_condition}) then becomes
\[
0=k_{B}k_{C}+k_{C}k_{A}(x^{2}+y^{2})+k_{B}k_{A}((x-1)^{2}+y^{2})\:,
\]
once developed and using the condition (\ref{eq:Scale_free rule})
we end up with 
\[
0=-k_{C}k_{A}-k_{C}k_{B}(x^{2}+y^{2})-2k_{B}k_{A}x\:,
\]
which leads to 
\begin{equation}
\left(x+\alpha\right)^{2}+y^{2}=R^{2}\:,\label{eq:Position_of_A}
\end{equation}
with, using the notation $\alpha=k_{A}/k_{C}$, $R^{2}=\alpha^{2}+\alpha+1>0$.
We always have a solution and a given shape of our triangle. In fact
only half of the circle correspond to expanding solutions, the other
half correspond to collapsing ones.

\subsection{Numerical study}

\begin{figure}
\begin{centering}
\includegraphics[width=7cm]{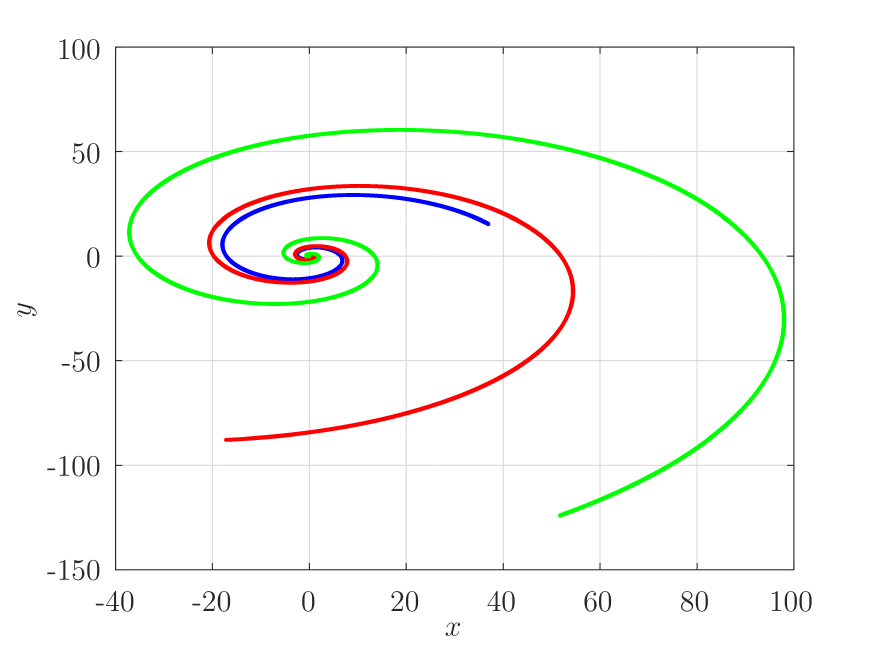}\\
\includegraphics[width=7cm]{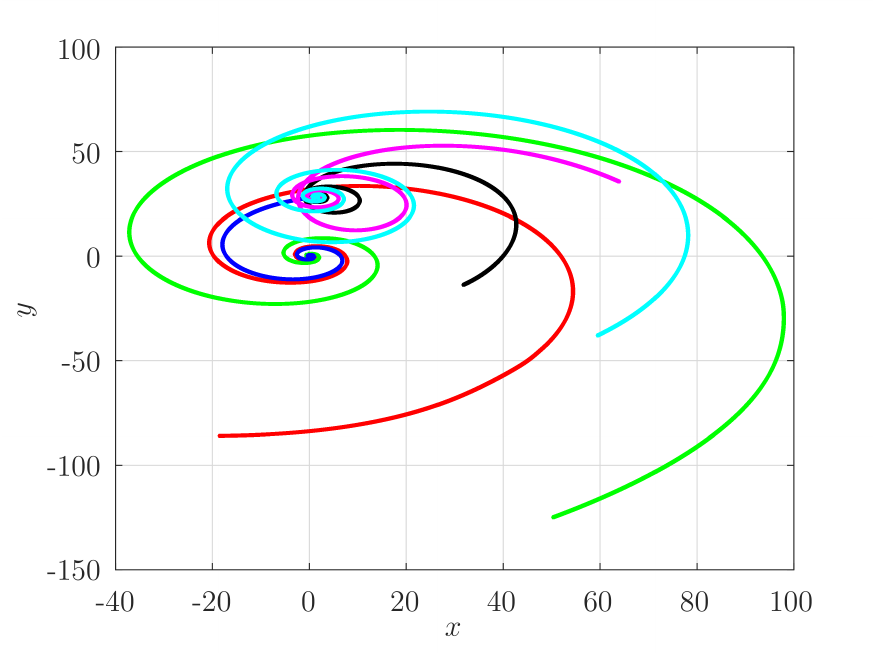}
\par\end{centering}
\caption{Top: Trajectories after one splitting up to time $T=20000$. After
the first division, the vortex strengths are $k_{1}=0.5924$ (red),
$k_{2}=0.7358$ (blue) and $k_{3}=-0.3282$ (green), what is striking
between the two figures is that the vortex $1$ and $3$ appear to
have the same trajectories in both cases.\protect \\
Bottom: Trajectories of vortices after 2 consecutive splitting up
to time $T=20000$. After the first division, the vortex strengths
are $k_{1}=0.5924$ (red), $k_{2}=0.7358$ (blue) and $k_{3}=-0.3282$
(green), the opposite values of Fig.~\ref{fig:Votrex_collapse}.
The second splitting occurs at time $T=10000$, the blue vortex is
divided and the $k_{2}'=0.637$ (black), $k_{4}=0.305$ (magenta)
and $k_{5}=-0.2063$ (cyan).\label{fig:Three-and-five-vortices}}
\end{figure}

\begin{figure}
\begin{centering}
\includegraphics[width=7cm]{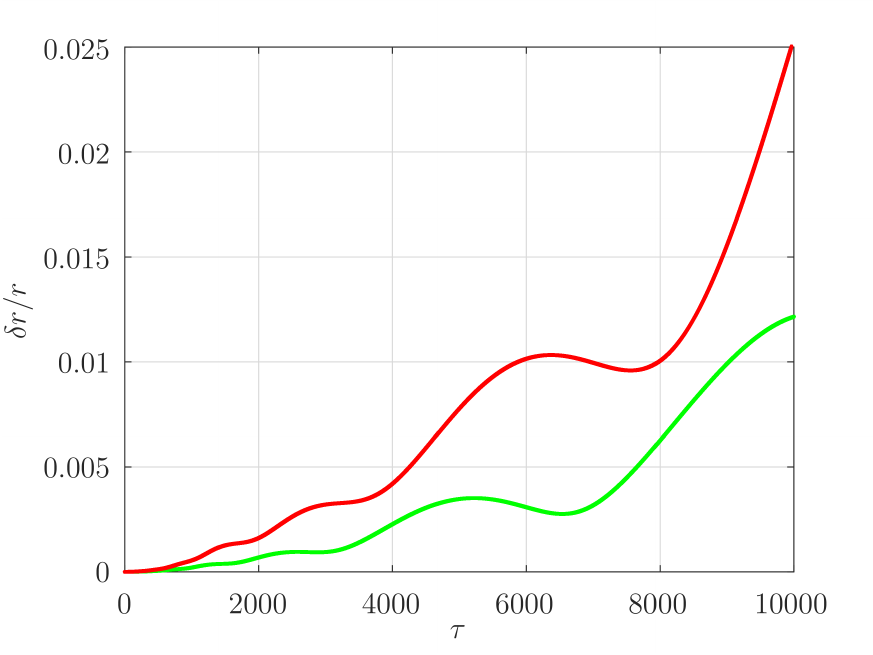}\\
\includegraphics[width=7cm]{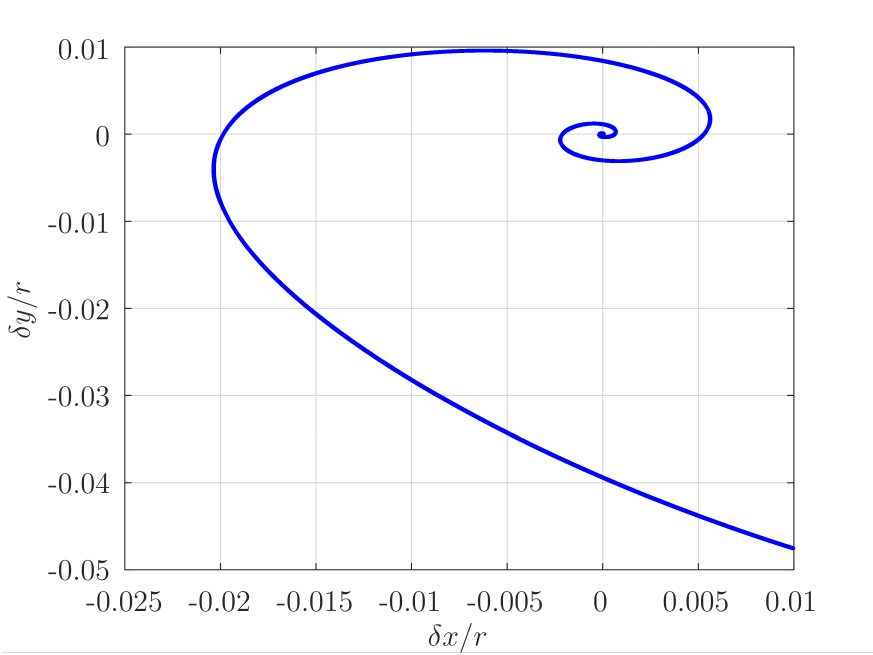}
\par\end{centering}
\caption{Top: Relative distance between the two trajectories of both vortices
that are not divided as a function of time. The relative error appears
to be growing but remains quite small even for large times, given
that the initial triangle after breakup is numerically set to $0.01$.
(the colors refer to the same vortices as in Fig.~\ref{fig:Three-and-five-vortices}).
The trajectories appear as almost unaffected by the splitting of the
third vortex.\protect \\
Bottom: Relative position of the center of vorticity of the three
newly created vortices with respect to the trajectory of the vortex
blue vortex in the top plot of Fig.~\ref{fig:Three-and-five-vortices}
. The center of vorticity of the three newly created vortices seems
with a very good approximation to follow the initial parent vortex
if it had not been broken up.\label{fig:Relative distances}}
\end{figure}

In order to study the dynamics numerically, we start with one vortex
of strength $k_{0}=1$. In order to split it into three vortices forming
a triangle $\hat{ABC}$, we first choose a random angle, which gives
us the values $k_{A}$, $k_{B}$, $k_{C}$. We then choose a second
angle to get the position of the point $A$ from Eq.~(\ref{eq:Position_of_A}),
(we make sure it corresponds to an expanding solution). After that
we center our triangle on the center of vorticity (which we fix to
be the point $(0,0)$), after that we choose again randomly an angle
and rotate the triangle around the center of vorticity, finally we
rescale the triangle to make it small (we set the largest side typically
to be $0.01$). Once the vortex is split we integrate numerically
the dynamics of the vortices, with a 6th-order simplectic scheme,
i.e Gauss-Legendre implicit scheme described in this paper\cite{McLachlan92},
with a time step of $\delta t=10^{-4}$. With a final total time of
the simulation $T_{f}=10000$ we recover the trajectories described
in Fig.~\ref{fig:Votrex_collapse} (we just change the signs of the
different vortex strengths to reverse the time). We then repeat the
previous operation, we pick one of the three vortices (our choice
is made the one with the largest absolute value for $k$), divide
it again in three, the only difference is that now the new triangle
is centered on the position of the vortex that was divided, and again
let the system evolve for another $T_{f}=10000$, the trajectories
of the initial three vortices for up to $T_{f}=20000$, and the ones
with three then five vortices are displayed in Fig.~\ref{fig:Three-and-five-vortices}.
 What is surprising is that the global dynamics of the five vortex
system, does not appear as chaotic, meaning that the trajectories
seem relatively smooth, what is even more surprising is that the trajectories
of the two vortices that have not been broken up, appear to be identical
after one of the three vortex is divided. A closer look at the deviation
from the original trajectories is provided in Fig.~\ref{fig:Relative distances}.
 This numerical analysis seems to indicate that indeed the unbroken
vortices trajectories are quasi unaffected by the splitting of the
third vortex, meaning that they pursue their trajectory as if the
third vortex was still there, moreover the center of vorticity of
the three newly created vortices appear to follow with a quite good
approximation the trajectory of the third vortex if it had not been
broken up.

This phenomenon which we dubbed vortex shadowing, triggered some real
surprise, indeed the breakup appears as ``invisible'' for the remaining
vortices, meaning that they are unable to see or ``measure'' the fact
that some singular and drastic event emerged. In order to check if
this phenomenon is intimately related to the presence of the singularity,
let us perform an analytical analysis.

\section{Analytical point of view}

In order to check whether two trajectories are identical one living
in a ten-dimensional phase space with 3 independent constant of the
motion, while the other one lives in a 6-dimensional phase space with
as well 3 independent constants, did not appear as an easy task. After
testing various different approaches with no success we settled for
one that is rather no completely satisfying for a continuous Hamiltonian
flow, but would be more conclusive for a discrete map, or a numerical
integrator, and settled for reasoning by induction.

\subsection{global setting}

We shall assume that we already have our three point vortex system,
each vortex located in a given point, which we note $A$, $B$ and
$C$. The vortex located in $A$ being the one that will be divided.
In order to start we will concentrate on the trajectory of the vortex
located in point $B$. We note $B_{3}$ the vortex in the $3-$point
vortex system and $B_{5}$ the vortex in the $5-$point vortex system,
note as well that in the following the indices will not always be
written when no confusion is possible. We will assume that at a given
time $t_{0}$ the equality $B_{3}=B_{5}$ holds and it is the same
for the point $C$, and that as well both the velocities

\begin{equation}
\boldsymbol{v}_{B_{3}}=\frac{1}{4\pi}\hat{\boldsymbol{e}}_{z}\times\left(k_{C}\frac{\boldsymbol{BC}}{BC^{2}}+k_{A}\frac{\boldsymbol{BA}}{BA^{2}}\right)\:,\label{eq:vB3}
\end{equation}
and 
\begin{equation}
\boldsymbol{v}_{B_{5}}=\frac{1}{4\pi}\hat{\boldsymbol{e}}_{z}\times\left(k_{C}\frac{\boldsymbol{BC}}{BC^{2}}+\sum_{I}k_{I}\frac{\boldsymbol{BI}}{BI^{2}}\right)\label{eq:vB5}
\end{equation}
are equal at the same time $t_{0}$ (note that this assumption is
true when the three vortices are just created and the point $A$ becomes
a ``three-point''). In Eq.~(\ref{eq:vB5}), the points $I$ correspond
to the positions of the three newly created vortices and $k_{I}$
are their respective strengths, $\hat{\boldsymbol{e}}_{z}$ is the
unit vector perpendicular, to the plane where the vortex move, $\times$
represents the cross product and bold characters represent vectors.
In the following in order to be more compact we use the notion
\begin{equation}
\boldsymbol{u}_{\perp}=\hat{\boldsymbol{e}}_{z}\times\boldsymbol{u}\:.\label{eq:definition_de_perp}
\end{equation}
It is also useful to write the speed of one of the newly created vortices
(we will omit the $4\pi$ from now on, as they cancel out in our calculation)
\begin{equation}
\boldsymbol{v}_{I}=k_{B}\frac{\boldsymbol{IB}_{\perp}}{IB^{2}}+k_{C}\frac{\boldsymbol{IC}_{\perp}}{IC^{2}}+\sum_{J\ne I}k_{J}\frac{\boldsymbol{IJ}_{\perp}}{IJ^{2}}\:.\label{eq:vI_definition}
\end{equation}

Our induction reasoning is corresponds to the following statement,
assuming the above is true, if we find that both accelerations $\boldsymbol{a}_{B_{3}}$
and $\boldsymbol{a}_{B_{5}}$ are equal, using some kind of Euler
scheme, and deduce from it that the statement will be as well correct
at time $t_{0}+\delta t$, and then reiterate the steps. Of course
since the flow is continuous, this does correspond to an exact result,
but should anyhow hint at least for finite time that the numerically
observed phenomena are not a numerical artifact.

The acceleration $\boldsymbol{a}_{B_{3}}=d\boldsymbol{v}_{B_{3}}/dt$
gives 
\begin{equation}
\boldsymbol{a}_{B_{3}}=\left[k_{C}\left(\frac{\boldsymbol{v}_{c}-\boldsymbol{v}_{B}}{BC^{2}}-2\frac{\boldsymbol{BC}\cdot\left(\boldsymbol{v}_{c}-\boldsymbol{v}_{B}\right)}{BC^{4}}\boldsymbol{BC}\right)+C\leftrightarrow A\right]_{\perp}\:,\label{eq:accelerationB3}
\end{equation}
where the underscore $\perp$ denotes the cross-product with $\hat{\boldsymbol{e}}_{z}$,
and $\cdot$ refers to the simple scalar product. In order to make
the expressions somewhat more compact, we introduce the operator $S_{\boldsymbol{u}\boldsymbol{_{\perp}}}$
as follows
\begin{equation}
S_{\boldsymbol{u}\boldsymbol{_{\perp}}}(\boldsymbol{v})=\boldsymbol{v}-2\frac{\boldsymbol{v}\cdot\boldsymbol{u}}{u^{2}}\boldsymbol{u}\:,\label{eq:Definition de S_u}
\end{equation}
which to given a vector $\boldsymbol{v}$ associates a new vector
$S_{\boldsymbol{u}\boldsymbol{_{\perp}}}(\boldsymbol{v})$. We can
notice that the result of this operation is the symmetric of the vector
$\boldsymbol{v}$ with respect to the direction perpendicular to the
vector $\boldsymbol{u}$. We may as well notice that this operator
(\ref{eq:Definition de S_u}) is linear in $\boldsymbol{v}$, but
not in $\boldsymbol{u}$. Using this formalism we can write the acceleration
of $B$ in the $5-$vortex system as 
\begin{equation}
\boldsymbol{a}_{B_{5}}=\left[k_{C}S_{\boldsymbol{BC}_{\perp}}\left(\frac{\boldsymbol{v}_{c}-\boldsymbol{v}_{B}}{BC^{2}}\right)+\sum_{I}k_{I}S_{\boldsymbol{BI}_{\perp}}\left(\frac{\boldsymbol{v}_{I}-\boldsymbol{v}_{B}}{BI^{2}}\right)\right]_{\perp}\:.\label{eq:accelerationB5}
\end{equation}
The condition of the equality of both accelerations $\boldsymbol{a}_{B_{5}}=\boldsymbol{a}_{B_{3}}$
leads to the condition:
\begin{equation}
k_{A}S_{\boldsymbol{BA}_{\perp}}\left(\frac{\boldsymbol{v}_{A}-\boldsymbol{v}_{B}}{BA^{2}}\right)=\sum_{I}k_{I}S_{\boldsymbol{BI}_{\perp}}\left(\frac{\boldsymbol{v}_{I}-\boldsymbol{v}_{B}}{BI^{2}}\right)\:,\label{eq:Condition_non_perturbee}
\end{equation}
that needs to be satisfied. We were unfortunately not able to verify
it by a direct computation and insertion of equations of the type
(\ref{eq:vB5}). However, since the system is scale invariant, it
is possible to imagine that we rescale our system (and thus rescale
as well our time), so that our newly formed triangle of vortices is
quite small with respect to the other typical distances of the problem,
we will therefore try to verify Eq.~(\ref{eq:Condition_non_perturbee})
by performing a perturbative expansion of it.

\subsection{Perturbative expansion}

As mentioned in order to proceed with a perturbative expansion, we
assume the triangle formed by the new vortices is small so that the
distances $AI$, and $IJ$ are small, of characteristic scale $\varepsilon$
with respect to the other typical length (for instance $CA$ or $BA$).
It is important to note that in this setting, the speeds $\boldsymbol{v}_{I}$
contain terms of order $1/\varepsilon$.

\subsubsection{Preliminary remarks}

In order to anticipate the expansion, we start to list some properties
of the operator $S_{\boldsymbol{u}\boldsymbol{_{\perp}}}$ defined
in Eq.~(\ref{eq:Definition de S_u}), and some others that will be
useful (see Appendix~\ref{sec:Appendice_propriete_de_S} for more
details).

Let $R_{\theta}$ be a rotation of an angle $\theta$ in the plane,
we have 
\begin{equation}
S_{R_{\theta}\boldsymbol{u}\boldsymbol{_{\perp}}}=R_{2\theta}S_{\boldsymbol{u}\boldsymbol{_{\perp}}}\:,\label{eq:Propriete_rotation_de_S}
\end{equation}
at least when $\theta<\pi/2$. We can also find that
\begin{equation}
S_{\boldsymbol{u}\boldsymbol{_{\perp}}}(\hat{\boldsymbol{e}}_{z}\times\boldsymbol{v})=-\hat{\boldsymbol{e}}_{z}\times S_{\boldsymbol{u}\boldsymbol{_{\perp}}}(\boldsymbol{v})\:,\label{eq:Propriete_prod_vect_de_S}
\end{equation}
 and may as well notice that 
\begin{equation}
\boldsymbol{u}_{\perp}=\hat{\boldsymbol{e}}_{z}\times\boldsymbol{u}=R_{\pi/2}\boldsymbol{u}\:.\label{eq:definition_u_perp}
\end{equation}
.

With this in mind, we can start the perturbative expansion of Eq.~(\ref{eq:Condition_non_perturbee}),
and identify at each order in $\varepsilon$ both side of the equality.

\subsubsection{Terms of order $1/\varepsilon$}

In the left hand side of Eq.~(\ref{eq:Condition_non_perturbee})
there are no terms of order $1/\varepsilon$, while in the right hand
side only the close vortices contribute, we have then
\begin{equation}
\boldsymbol{0}=\sum_{I\ne J}\frac{k_{I}k_{J}}{IJ^{2}}S_{\boldsymbol{BI}_{\perp}}\left(\frac{\boldsymbol{IJ}_{\perp}}{BI^{2}}\right)\:,\label{eq:Perturb_Eq_1_sur_eps_1}
\end{equation}
here it appears to be useful to introduce the notation 
\begin{equation}
\alpha_{IJ}=\frac{k_{I}k_{J}}{IJ^{2}}\:,\label{eq:definition_alpha_ij}
\end{equation}
 which is symmetric in $I$ and $J$. At this order, all corrections
of order $\varepsilon$ are negligible, hence we may identify $\boldsymbol{BI}$
with $\boldsymbol{BA}$, and using the linear property of the operator
$S_{\boldsymbol{u}\boldsymbol{_{\perp}}}$ we obtain
\begin{equation}
\boldsymbol{0}=S_{\boldsymbol{BA}_{\perp}}\left(\sum_{I\ne J}\alpha_{IJ}\frac{\boldsymbol{IJ}_{\perp}}{BA^{2}}\right)\:,\label{eq:Perturb_Eq_1_sur_eps_2}
\end{equation}
and since $S_{\boldsymbol{u}\boldsymbol{_{\perp}}}$ is some kind
of symmetry, its kernel reduces to the null vector, and since the
cross product with $\hat{\boldsymbol{e}}_{z}$ is a rotation, it is
as well invertible, Eq.~(\ref{eq:Condition_non_perturbee}) reduces
to this order to 
\begin{equation}
\boldsymbol{0}=\frac{1}{BA^{2}}\sum_{I\ne J}\alpha_{IJ}\boldsymbol{IJ}\:.\label{eq:Perturb_Eq_1_sur_eps_final}
\end{equation}
In Eq.~(\ref{eq:Perturb_Eq_1_sur_eps_final}) the right side is antisymmetric
by the exchange in $I\leftrightarrow J$, it is thus equal to its
opposite, and we can conclude that Eq.~(\ref{eq:Condition_non_perturbee})
is verified at order $1/\varepsilon$. We now will move to the next
order and compute terms of order $0$, these comprise simple order
$0$ terms, but as well terms mixing orders $1/\varepsilon$ and $\varepsilon$.
We assume that these are independent, and treat them separately in
what follows.

\subsubsection{Simple terms of order $\varepsilon^{0}$}

To obtain the simple terms of order, just means discard the terms
of order $1/\varepsilon$, and consider that all $I$ points are actually
identical to $A$. This means that $S_{\boldsymbol{BI_{\perp}}}=S_{\boldsymbol{BA_{\perp}}}$,
and Eq.~(\ref{eq:Condition_non_perturbee}) reduced to these approximations
leads to
\begin{equation}
k_{A}\frac{\boldsymbol{v}_{A}-\boldsymbol{v}_{B}}{BA^{2}}=\sum_{I}k_{I}\frac{\boldsymbol{v}^{(0)}_{I}-\boldsymbol{v}_{B}}{BA^{2}}\:,\label{eq:Condition_non_perturbee-1}
\end{equation}
 where we retain only order $0$ terms in $\boldsymbol{v}_{I}$ ,
i.e we retain only the influence of vortex $B$ and $C$ in Eq.~(\ref{eq:vI_definition}).
This leads to 
\begin{equation}
k_{A}\boldsymbol{v}_{A}=\sum_{I}k_{I}\boldsymbol{v}^{(0)}_{I}\:,\label{eq:Condition_non_perturbee-1-1}
\end{equation}
and we find a first condition, that confirms what was observed numerically
in Fig.~\ref{fig:Relative distances}, i.e that\emph{ the motion
of $A$ is identical to the center of vorticity of the three new vortices}.
Let us now move on to the second term of order $0$.

\subsubsection{Terms of order $\varepsilon^{0}=\varepsilon\times1/\varepsilon$}

In the development we also have terms of order $0$ coming from the
product of terms of order $1/\varepsilon$ multiplied by terms of
order $\varepsilon$. As stated, the only contributions of order $1/\varepsilon$
comes from the interaction between new vortices. To obtain these terms
we then rewrite what is left of Eq.~(\ref{eq:Condition_non_perturbee})
as 
\begin{equation}
\boldsymbol{0}=\sum_{I\ne J}\alpha_{IJ}\left(S_{\boldsymbol{BI}_{\perp}}\left(\frac{\boldsymbol{IJ}_{\perp}}{BI^{2}}\right)-S_{\boldsymbol{BA}_{\perp}}\left(\frac{\boldsymbol{IJ}_{\perp}}{BA^{2}}\right)\right)\:,\label{eq:Condition_non_perturbee-2}
\end{equation}
which just means that we subtracted in Eq.~(\ref{eq:Condition_non_perturbee})
the terms of order $1/\varepsilon$ and the simple terms of order
$0$. To move on we add and remove the same term, and we end up with
the following
\begin{align}
\boldsymbol{0} & =\sum_{I\ne J}\alpha_{IJ}\left(S_{\boldsymbol{BI}_{\perp}}\left(\frac{\boldsymbol{IJ}_{\perp}}{BI^{2}}\right)-S_{\boldsymbol{BI}_{\perp}}\left(\frac{\boldsymbol{IJ}_{\perp}}{BA^{2}}\right)\right)\label{eq:Condition_non_perturbee-2-1}\\
+ & \sum_{I\ne J}\alpha_{IJ}\left(S_{\boldsymbol{BI}_{\perp}}\left(\frac{\boldsymbol{IJ}_{\perp}}{BA^{2}}\right)-S_{\boldsymbol{BA}_{\perp}}\left(\frac{\boldsymbol{IJ}_{\perp}}{BA^{2}}\right)\right)\:,
\end{align}
after removing the $\perp$ indices using the relation~(\ref{eq:Propriete_prod_vect_de_S}),
and the linear property of $S$, we obtain 
\begin{align}
\boldsymbol{0} & =\sum_{I\ne J}\alpha_{IJ}S_{\boldsymbol{BI}_{\perp}}\left(\boldsymbol{IJ}\right)\left(\frac{1}{BI^{2}}-\frac{1}{BA^{2}}\right)\label{eq:Condition_non_perturbee-2-1-1}\\
+ & \sum_{I\ne J}\alpha_{IJ}\frac{1}{BA^{2}}\left(S_{\boldsymbol{BI}_{\perp}}-S_{\boldsymbol{BA}_{\perp}}\right)\left(\boldsymbol{IJ}\right)\:.
\end{align}
Let us analyze those terms separately. In the first term the order
$\varepsilon$ comes from 
\[
\left(\frac{1}{BI^{2}}-\frac{1}{BA^{2}}\right)\approx2\frac{\boldsymbol{BA}\cdot\boldsymbol{AI}}{BA^{4}}\:,
\]
and the term of order $1/\varepsilon$ from 
\[
\alpha_{IJ}S_{\boldsymbol{BI}_{\perp}}\left(\boldsymbol{IJ}\right)\approx\alpha_{IJ}S_{\boldsymbol{BA}_{\perp}}\left(\boldsymbol{IJ}\right)\:,
\]
as we only keep the main part of it.

For the second term, let us note $\varepsilon_{I}$ the angle between
$\boldsymbol{BI}$ and $\boldsymbol{BA}$, since the operator $S_{\boldsymbol{u}_{\perp}}$
only depends on the direction of $\boldsymbol{u}$ and not its norm
. We have 
\begin{align*}
S_{\boldsymbol{BI}_{\perp}}-S_{\boldsymbol{BA}_{\perp}} & =S_{R(\varepsilon_{i})\boldsymbol{BA}_{\perp}}-S_{\boldsymbol{BA}_{\perp}}\\
 & =(R(2\varepsilon_{i})-\mathbb{\mathbb{I}})S_{\boldsymbol{BA}_{\perp}}\\
 & \approx2\varepsilon_{i}\hat{\boldsymbol{e}}_{z}\times S_{\boldsymbol{BA}_{\perp}}\:.
\end{align*}
Using Eq.~(\ref{eq:Propriete_prod_vect_de_S}), this implies that
\[
\left(S_{\boldsymbol{BI}_{\perp}}-S_{\boldsymbol{BA}_{\perp}}\right)\left(\boldsymbol{IJ}\right)\approx-2\varepsilon_{i}S_{\boldsymbol{BA}_{\perp}}(\boldsymbol{IJ}_{\perp})\:.
\]
We now use the fact that keeping things at first order we have $BA=BI$,
and using a definition of the cross product we have 
\[
\varepsilon_{i}\approx\sin(\varepsilon_{i})\approx\frac{\boldsymbol{BI}\times\boldsymbol{BA}}{BA^{2}}\cdot\hat{\boldsymbol{e}}_{Z}=\frac{\boldsymbol{AI}\times\boldsymbol{BA}}{BA^{2}}\cdot\hat{\boldsymbol{e}}_{Z}\:.
\]
The total term~(\ref{eq:Condition_non_perturbee-2-1-1}) reduces
then to 
\[
2S_{\boldsymbol{BA}_{\perp}}\left(\sum_{I\ne J}\alpha_{IJ}\left(\frac{\boldsymbol{BA}\cdot\boldsymbol{AI}}{BA^{4}}\boldsymbol{IJ}-\frac{(\boldsymbol{AI}\times\boldsymbol{BA})\cdot\hat{\boldsymbol{e}}_{Z}}{BA^{4}}\boldsymbol{IJ}_{\perp}\right)\right)\:.
\]
The equation~(\ref{eq:Condition_non_perturbee-2-1-1}) is then satisfied
at this order if 
\[
\boldsymbol{0}=\sum_{I\ne J}\alpha_{IJ}\left((\boldsymbol{BA}\cdot\boldsymbol{AI})\boldsymbol{IJ}-\left((\boldsymbol{AI}\times\boldsymbol{BA})\cdot\hat{\boldsymbol{e}}_{Z}\right)\boldsymbol{IJ}_{\perp}\right)\:,
\]
to simplify a little more, we exchange $I$ and $J$ in the previous
sum, add the two identical equations to remove its antisymmetric component
and we obtain
\[
\boldsymbol{0}=\sum_{I\ne J}\alpha_{IJ}\left((\boldsymbol{BA}\cdot\boldsymbol{IJ})\boldsymbol{IJ}-\left((\boldsymbol{IJ}\times\boldsymbol{BA})\cdot\hat{\boldsymbol{e}}_{Z}\right)\boldsymbol{IJ}_{\perp}\right)\:.
\]
Since $(\boldsymbol{IJ}\times\boldsymbol{BA})\cdot\hat{\boldsymbol{e}_{Z}}=-\boldsymbol{BA}\cdot\boldsymbol{IJ}_{\perp}$
we obtain
\[
\boldsymbol{0}=\sum_{I\ne J}k_{I}k_{J}\left(\left(\boldsymbol{BA}\cdot\frac{\boldsymbol{IJ}}{IJ}\right)\frac{\boldsymbol{IJ}}{IJ}+\left(\boldsymbol{BA}\cdot\frac{\boldsymbol{IJ}_{\perp}}{IJ}\right)\frac{\boldsymbol{IJ}_{\perp}}{IJ}\right)\:,
\]
which reduces to 
\[
\boldsymbol{0}=\left(\sum_{I\ne J}k_{I}k_{J}\right)\boldsymbol{BA}\:.
\]
We recover the conditions necessary for the singularity (\ref{eq:scale_invariance_gen}).

The same computations could be performed for $C$ as long as it is
as well far from the three new vortices.

As a conclusion for this analytical expansion, we may say that the
phenomenon of vortex shadowing that was observed numerically is indeed
a real phenomena, at least for times up to $1/\varepsilon$, since
we did not carry our expansion to higher orders, and as long as the
points remain sufficiently far away, using the scale invariance and
rescaling the system, we could even imagine that this could be valid
for much larger times, as is observed (the typical numerical scale
of our new triangle is set to $0.01$ in our simulations, corresponding
to a relative $\varepsilon\approx0.001$, inducing a validity up to
$t\approx1000$, and our observation are convincing for at least one
order of magnitude more). And that a necessary condition for this
phenomenon to be observed is the scale invariance symmetry (\ref{eq:scale_invariance_gen})
of the Hamiltonian of the system.

\section{Consequences and discussions}

\begin{figure}
\begin{centering}
\includegraphics[width=7cm]{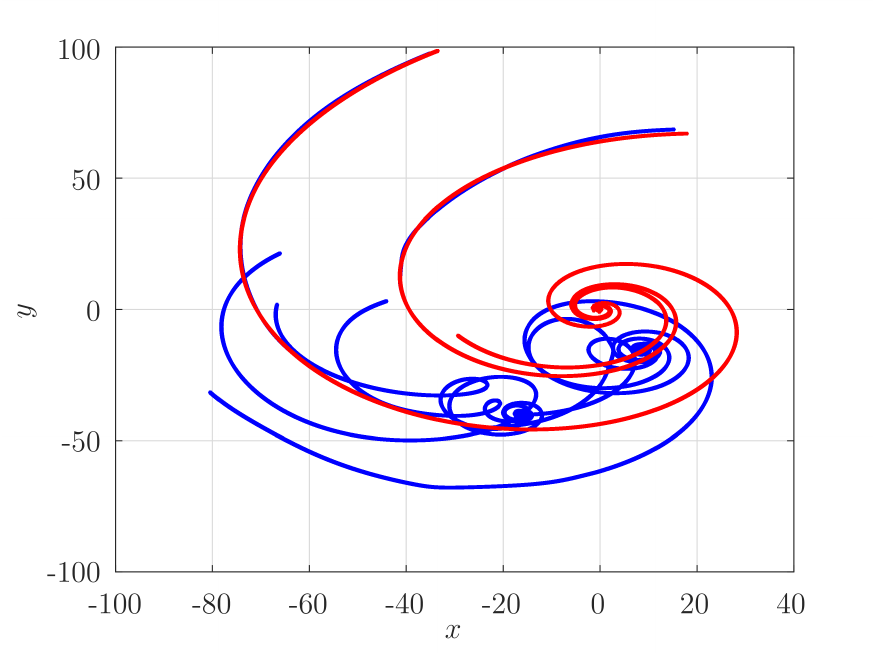}\\
\includegraphics[width=7cm]{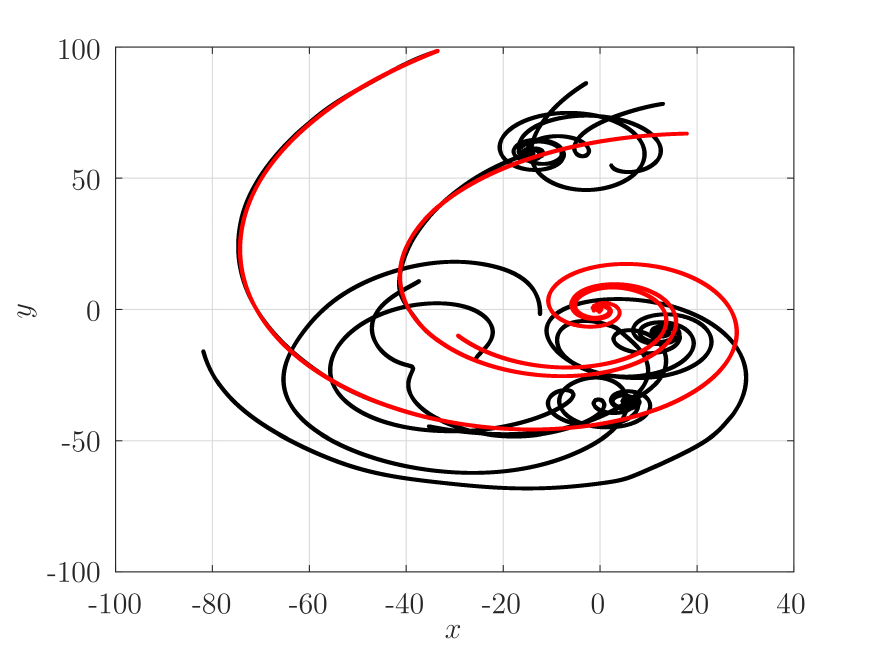}
\par\end{centering}
\caption{Top: Comparison of trajectories up to time $T=12000$, in red trajectories
after an initial split. In blue, one of the three initial vortice
is split in three at time $t_{1}=4000$, and consecutively one of
the created vortices is again split in three at time $t_{2}=8000$,
and we end up with 7 vortices. One can notice that the trajectories
of the initial undivided vortices remain very similar. \protect \\
Bottom: Comparison of trajectories up to time $T=12000$, in red trajectories
after an initial split. In black, one of the three initial vortice
is split in three at time $t_{1}=3000$, consecutively one of the
created vortices is again split in three at time $t_{2}=6000$, and
finally one of the initially created vortices is split at time $t_{3}=9000$,
we end up with a 9-vortex system. We notice that the trajectories
of the initially created and remaining undivided vortex remain very
similar in both cases. \label{fig:more_3_splits}}
\end{figure}

Looking back at the analytical results obtained in the previous section,
we notice that the emerging condition is only imposed on the three
new vortices that are formed, and that if the condition (\ref{eq:scale_invariance_gen})
is satisfied, than the speed/trajectory of the non divided vortices
is not changing, provided $\varepsilon$ is sufficiently small. If
this is correct, this means that if we perform successive divisions
from for instance three to five vortices, then five to seven or event
seven to nine, if all vortices are sufficiently far from each other,
a vortex that is undivided will keep the same trajectory for a given
time. Results of this scenario are confirmed in Fig.~\ref{fig:more_3_splits}.
 
\begin{figure}
\begin{centering}
\includegraphics[width=7cm]{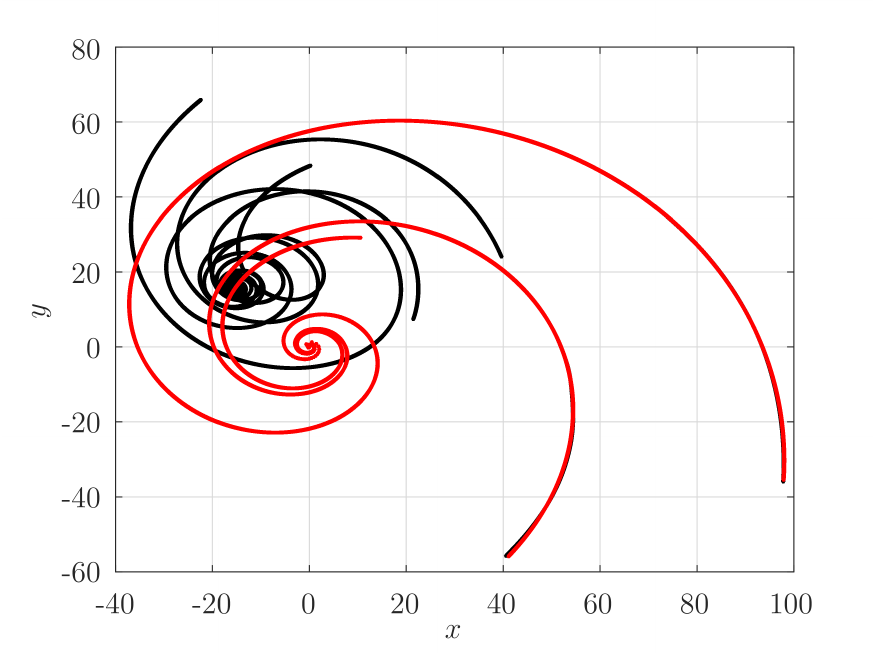}\\
\includegraphics[width=7cm]{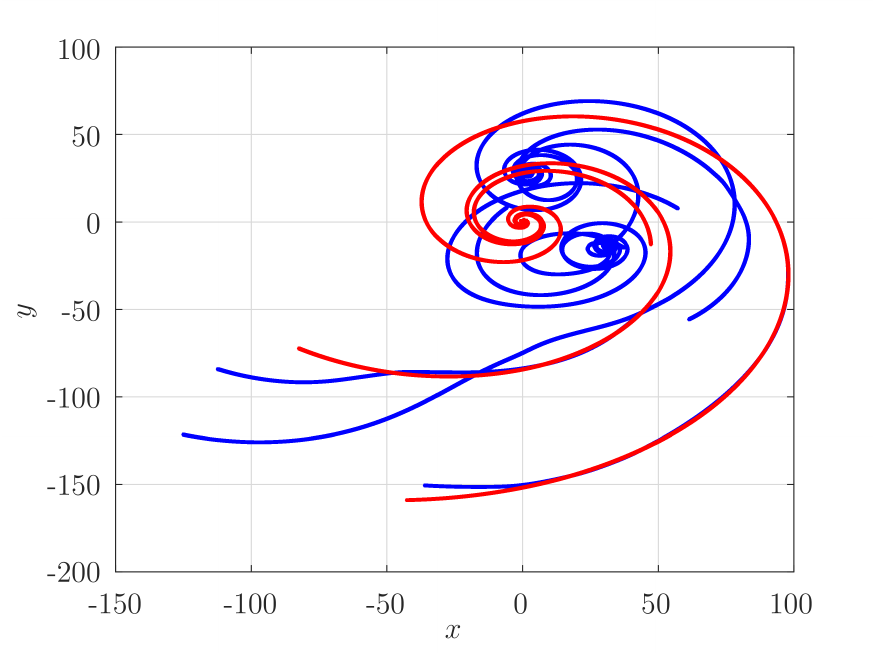}
\par\end{centering}
\caption{Top: Comparison of trajectories up to time $T=12000$, in red trajectories
after an initial split. In black, one of the three initial vortice
is split in four vortices at time $t_{1}=6000$ and we end up with
6 vortices. One can notice that the trajectories of the initial undivided
vortices remain very similar. \protect \\
Bottom: Comparison of trajectories up to time $T=30000$, in red trajectories
after an initial split. In blue, one of the three initial vortice
is split in three at time $t_{1}=10000$, consecutively one of the
created vortices is again split in three at time $t_{2}=20000$. We
notice that the trajectories of the initially created voritces diverge
from their initial ones after one of the newly created vortices comes
``close'' to one of them and breaks the assumption $\varepsilon\ll1$
even after rescaling used in the perturbative expansion.. \label{fig:Split_in_4_and_error_with_split_in_3}}
\end{figure}
 It is thus clear that in a setting compatible with scale invariance
an isolated vortex that is not divided can not perceive, i.e measure,
the fact that other vortices are divided, and continues on its trajectory
as if nothing happened.

Another consequence of our analytical result correspond to the fact
that the computation even when it was thought for a singularity corresponding
to reverse three vortex collapse course, is actually independent on
the number of vortices generated during the singularity, meaning if
we had more than three vortices the calculations would be identical
and we just would have more terms in the sums over $I$ or over $I$
and $J$. In order to verify if this indeed the case we considered
a numerical experiment in which after a vortex is split in three,
a second splitting of one vortex occurs but this time using a singularity
generating four vortices. The results of this simulation are described
in Fig.~\ref{fig:Split_in_4_and_error_with_split_in_3}, and indeed
we clearly see that the trajectories of the undivided vortices are
not affected wether the third vortex is split in four new vortices
or not, like what had been observed when it was or not divided in
three. The details on how to construct an initial condition reflecting
this 4 vortex singularity is given in the Appendix~\ref{sec:Appendice_Singularite_4}.
Even though we did not perform a numerical test, we may as well anticipate
that this shall as well be the case when we use singularities generating
even more than four vortices.

Finally, it is also important to notice, that this phenomenon is not
necessarily eternal, and when one of the newly created vortices comes
sufficiently close to one of the initial vortices, which we may consider
as some type of collision, the hypothesis that $\varepsilon$ is small
breaks. This is illustrated as well in Fig.~\ref{fig:Split_in_4_and_error_with_split_in_3}.

To conclude this study, and just considering 3-vortex divisions. Let
us insist on our findings the remaining vortices do not have a change
in their trajectory, and the newly created system of three vortices
has its center of vorticity following the trajectory of the ``phantom''
of the destroyed vortex as if it was still present. Since the initial
system is integrable the trajectories are exactly known, meaning that
in the ten dimensional phase space, already six dimensions are known,
moreover we still have a global conservation of energy, of the global
center of vorticity and as well angular momentum, we may thus expect
to fall on a one-dimensional phase space and thus somehow remain integrable
and be able to analytically predict the course of the 5 vortices,
moving then on to seven and continuing the divisions thereafter we
will have exactly the same scenario. This means that we could be able
to have access to the dynamics of infinitely many vortices, and going
to some continuous limit to an evolution of a vorticity distribution.
One of the foreseeable difficulty in achieving this will most probably
being able to control the $\varepsilon$ term and avoid ``collisions'',
but given the liberty in the choice of vortex strengths (on random
angle), as well on the ``triangle'' (a tri-point actually) shape
(one random angle), and on its orientation with respect to the remaining
vortices (on random angle as well), this may not be hopeless.

Finally, we would like to emphasize on what this phenomenon teaches
us, if we were to model and deduce physical laws and only had the
interaction between vortices to do so, should we be in a situation
with a conformal invariance and analyzing our own motion we may conclude
on the presence of two other vortices, while there could actually
be many more present, leading to potential unexpected collisions,
and a thus a false representation of our physical world. With this
in mind, we could question if the quest of unifying all physical interactions
in a unique law that could lead to singularities, is actually a good
one, and if having more interactions is not in the end a good thing,
that keep us discovering the world for what it is and not what we
would like it to be.

\begin{acknowledgments}
The authors would like to acknowledge support from the M25P01 ECOS Nord program.

X.L and E.U acknowledge as well support from the IEA program of the CNRS that allowed to lay out the initial work.

X. L would like to mention that this work has been carried out within the framework
of the EUROfusion Consortium, funded by the European
Union via the Euratom Research and Training Programme
(Grant Agreement No 101052200 — EUROfusion). Views
and opinions expressed are however those of the author(s)
only and do not necessarily reflect those of the European
Union or the European Commission. Neither the Euro-
pean Union nor the European Commission can be held
responsible for them.

\end{acknowledgments}

\appendix

\section{Splitting a vortex in 4\label{sec:Appendice_Singularite_4}}

In order to test the consequence of a division in four different vortices,
we describe after how we choose our conditions to perform the simulation.
We did not carry a full general solution, and settled for a simple
usable subset of conditions, in order to compute the results displayed
in Fig.~\ref{fig:Split_in_4_and_error_with_split_in_3}.

\subsection{Vortex strength}

We first find the conditions for the strengths and we just use Eqs
(\ref{eq:Vorticity conservation} and \ref{eq:Scale_free rule}) just
extended to four vortices. We first settled for a simple case where
we have $(1,1,1,-1)$, but if two opposite vortices come close to
each other and form locally a dipole, they quickly leave towards infinity
and is not really interesting to test our problem. So we ended up
looking for a solution of the form $(1,2,k,-1/2)$. Using the scale
invariant condition~(\ref{eq:Scale_free rule}) we end up with
\[
\frac{1}{2}+\frac{5}{2}k=0\:,
\]
i.e $k=-1/5$. With this setting the total vorticity becomes $\sum k_{i}=23/10$,
and we will have to rescale all of them to reflect the strength of
the divided vortex (\ref{eq:Vorticity conservation}). With this setting
not dipole is expected and a longer interaction between the vortices
is expected.

\subsection{Geometric condition}

The geometric condition follows from Eq.(\ref{eq:ang_momentum_condition}),
noting $r_{ij}$ the distance between the four newly created vortices
and using the selected vortex strengths, this equation becomes
\begin{align*}
2r^{2}_{12}-\frac{1}{5}r^{2}_{13}-\frac{1}{2}r^{2}_{14}-\frac{2}{3}r^{2}_{23}-r^{2}_{24}+\frac{1}{10}r^{2}_{34} & =0\:,
\end{align*}
which we rewrite
\begin{equation}
2r^{2}_{12}-\frac{1}{5}r^{2}_{13}-\frac{2}{3}r^{2}_{23}=\frac{1}{2}(r^{2}_{14}+2r^{2}_{24}+\frac{1}{5}r^{2}_{34})\:,\label{eq:Eq_pour_4_vortex}
\end{equation}
isolating vortex number 4 in one side.

Our strategy is then to follow what we did for three vortices, we
fix the coordinates of the first 3 vortices and look on a condition
for the fourth, contrary to the 3-vortex split we loose some generality
by fixing the third vortex position, but our goal is just to find
an example. The position of vortex 1 is set to $(0,0)$, the position
of vortex 2 is set to $(1,0)$, the position of vortex 3 is set to
$(0,1)$, creating an isosceles right triangle. We look now for the
coordinates $(x,y)$ of the fourth vortex, by using these values in
Eq.~(\ref{eq:Eq_pour_4_vortex}), this leads to 
\[
(r^{2}_{14}+2r^{2}_{24}+\frac{1}{5}r^{2}_{34})=2\:,
\]
which after some simple calculus leads to 
\[
(x_{4}-\frac{5}{7})^{2}+(y_{4}+\frac{1}{14})^{2}=\frac{115}{196}\:.
\]
Our fourth vortex lies on a circle of radius $R_{4}=\sqrt{115}/14$
centered on the point $(5/7,\,-1/14)$.

In the simulation carried in Fig.~\ref{fig:Split_in_4_and_error_with_split_in_3},
we used and angle $\pi/3$ on the circle, leading to a vortex position
$R_{4}(1/2,\,\sqrt{3}/2)-(5/7,\,-1/14)$ . We then rescaled the quadrilateral
so that the maximum distance between two points is $0.01$, and translated
it so to equal the position of the center of vorticity of the quadrilateral
to the position of the divided vortex. Finally we rescaled each vortex
strength to insure the conservation of vorticity.

\section{Properties of $S_{\boldsymbol{u}_{\perp}}$\label{sec:Appendice_propriete_de_S}}

\begin{figure}
\begin{centering}
\includegraphics[width=7cm]{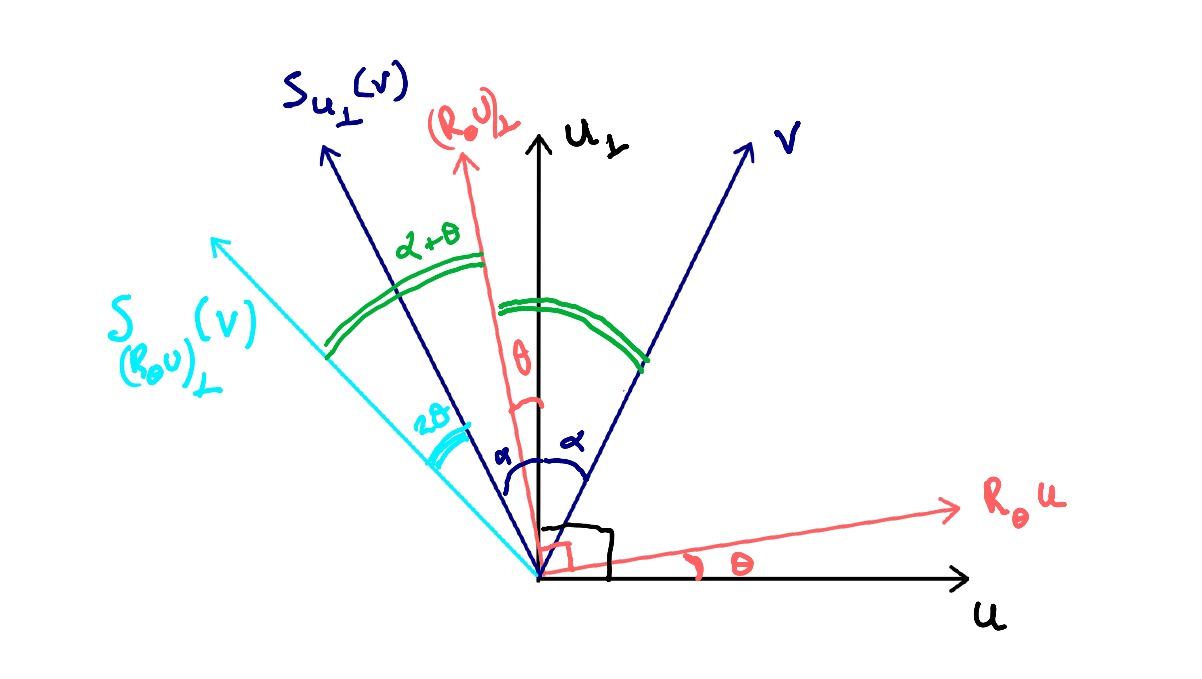}\\
\includegraphics[width=7cm]{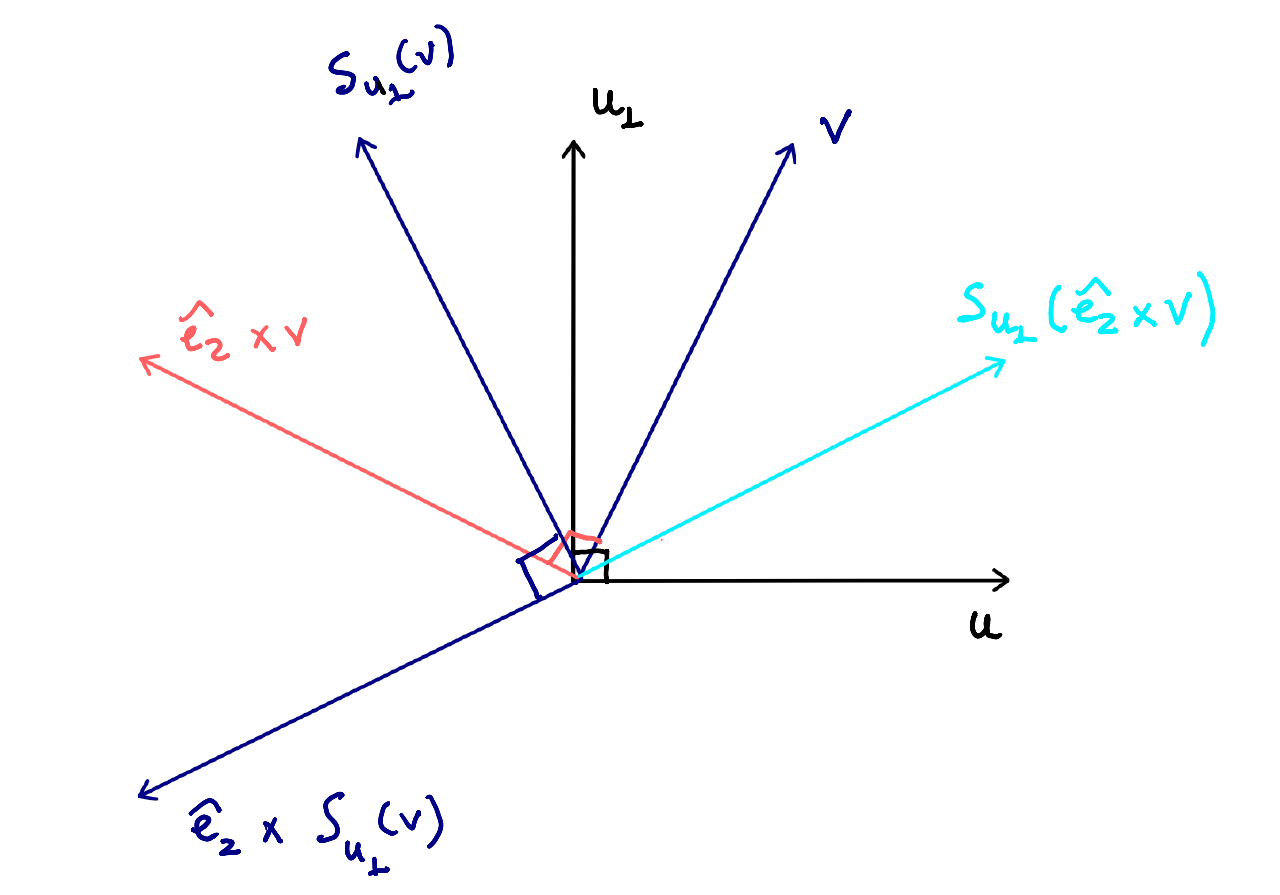}
\par\end{centering}
\caption{Top: Sketch to get the $S_{R_{\theta}\boldsymbol{u}\boldsymbol{_{\perp}}}=R_{2\theta}S_{\boldsymbol{u}\boldsymbol{_{\perp}}}$
property. \protect \\
Bottom: Sketch to get the $S_{\boldsymbol{u}\boldsymbol{_{\perp}}}(\hat{\boldsymbol{e}}_{z}\times\boldsymbol{v})=-\hat{\boldsymbol{e}}_{z}\times S_{\boldsymbol{u}\boldsymbol{_{\perp}}}(\boldsymbol{v})\:,$
property. \label{fig:Proprietes_de_S_u}}
\end{figure}

While the properties (\ref{eq:Propriete_rotation_de_S} and \ref{eq:Propriete_prod_vect_de_S})
of the operator $S_{\boldsymbol{u}_{\perp}}$, could be derived with
analytical computations, it seemed much simpler and intuitive to do
so geometrically with the sketches that are displayed in Fig.~\ref{fig:Proprietes_de_S_u}.

\bibliography{extracted}
\end{document}